\documentclass{aa}
\usepackage{psfig}

\begin{document}

\title{Dust emission and star formation toward a redshift 5.5 QSO}

\author{Frank Bertoldi\inst{1}
       \and Pierre Cox\inst{2}}

\offprints{F.~Bertoldi, bertoldi@mpifr.bonn-mpg.de}

\institute{Max-Planck-Institut f\"ur Radioastronomie, 
Auf dem H\"ugel 69, D-53121 Bonn, Germany
\and Institut d'Astrophysique Spatiale,  Universit\'e de Paris XI, 
F-91405 Orsay, France}

\date{Received date / Accepted date}

\titlerunning{Dust emission from RD0301}
\authorrunning{F. Bertoldi \& P. Cox}

\abstract{We report observations of the low-luminosity $z = 5.50$ quasar
  RD~J030117+002025 (RD0301 hereafter) at 250 GHz (1.20~mm) using the
  Max-Planck Millimeter Bolometer (MAMBO) array at the IRAM 30-meter
  telescope. The quasar was detected with a 1.2~mm flux density of $0.87
  \pm 0.20 \, \rm mJy$. The lack of detectable 1.4 GHz 
  radio emission indicates
  that the millimeter emission is of thermal nature, making RD0301 the
  most distant dust-emission source known. When matching a 50~K grey body
  thermal far-infrared (FIR) spectrum to the observed millimeter flux
  we imply a FIR luminosity $\approx \rm 4 \times 10^{12} \, L_{\odot}$,
  which is comparable to the quasar's optical luminosity.  If the FIR
  luminosity arises from massive star formation, the implied star
  formation rate would be $\sim 600\,\rm M_\odot yr^{-1}$, comparable
  to that of the starburst galaxies which dominate the average star
  formation and FIR emission in the early Universe.  The FIR luminosity
  of RD0301 is close to the average of that found in optically far more
  luminous high-redshift quasars.  The comparably high millimeter to optical
  brightness ratio of RD0301 is further evidence for that there is no
  strong correlation between the optical and millimeter brightness of
  high-redshift quasars, supporting the idea that in high-redshift
  quasars the dust is not heated by the AGN, but by starbursts.
   }

\maketitle

\sloppy

\section{Introduction}

Many high-redshift quasars (QSOs) have recently been found through wide
field imaging surveys, notably the Sloan Digital Sky Survey (SDSS;
Schneider et al. \cite{sch01}; Anderson et al.  \cite{and01}) and the
Digital Palomar Sky Survey (DPSS; Kennefick et al.
\cite{ken95a},\cite{ken95b}; Djorgovski et al. \cite{djo99}). Now over
250 QSOs are known with redshifts $z>3.6$, thirteen of which are at
$z>5$ (Fan et al. \cite{fan99}, \cite{fan00a}, \cite{fan00b}; Zheng et
al.  \cite{zhe00}; Stern et al.  \cite{ste00}; Sharp et al.
\cite{sha01}; Fan et al. \cite{fan01}) and one at $z>6$ (Fan et al.
\cite{fan01}).  All but one of the $z>5$ QSOs were found from magnitude
$i\approx 21.5$ limited surveys, limiting their implied rest frame blue
magnitudes to $M_{\rm B} < -26$.

In a 74 arcmin$^2$ field observed for the SPICES survey (Stern et al.
\cite{ste01}), Stern et al. (\cite{ste00}) discovered the redshift 5.50
QSO RD (i.e., ``R-band Dropout'') 
J030117+002025 (RD0301 hereafter), which has
an $i$-band AB magnitude of 23.4
and $M_{\rm B} \approx -23.4$ in an Einstein-de Sitter Universe
($H_0=50~ \rm km~s^{-1} Mpc^{-1}$, $q_0=0.5$), and $M_{\rm B} \approx
-24.0$ in a $\Lambda$-cosmology ($H_0=65\rm~km~s^{-1}~Mpc^{-1}$,
$\Omega_\Lambda=0.7$, $\Omega_{\rm m}=0.3$) which we adopt throughout
this paper.  RD0301 is significantly fainter than the typical QSO found
in the wide field SDSS and DPSS surveys.  The discovery of such a very
high redshift QSO was surprising because the surface density of such
objects, as implied by the intermediate-redshift QSO luminosity function
and redshift evolution models, is so low that the probability to find
one in the observed field is only $\sim 15\%$ (Stern et al.
\cite{ste00}).  The discovery may therefore imply that the luminosity
evolution of faint high-redshift QSOs is weaker than expected, and that
there could be a larger, optically faint QSO population at high
redshifts. This possibility is particularly interesting in terms of
which sources re-ionize the early Universe (cf., the recent
Gunn-Peterson trough results of Becker et al. [\cite{bec01}] and
Djorgovski et al.  [\cite{djo01}]).


Another important question in studies of the early Universe is how the
growth of massive black holes relates to the formation of early stellar
populations. A tight correlation between black hole masses and bulge
luminosities or velocity dispersions in local galaxies indicates that
their formation was closely related (Magorrian et al. \cite{mag98}).
The recent detection of strong thermal dust emission from many high-redshift
QSOs shows that vigorous star formation is coeval with black hole growth
(Carilli et al. \cite{car01a}; Omont et al. \cite{omo01}; Isaak et al.
\cite{isa01}).  Interestingly, the FIR luminosity of these QSOs does not
seem to be well correlated with their optical luminosity in the studied
range of optical luminosities, $M_{\rm B} = -26$ to $-29.5$.  A similar
result was previously found also in lower redshift, lower optical
luminosity QSOs (Sanders et al. \cite{san89}; Chini et al.
\cite{chi89}).  Such a lack of correlation might suggest that the warm
dust is not heated by the QSO. Instead, star formation and black hole
accretion could have had a common cause, which might be the infall
of material toward the center of the host galaxy.

The QSOs observed in the millimeter and submillimeter surveys are the
optically most luminous objects, and their implied optical luminosities
are on average ten times higher than their FIR luminosities, so that
dust heating by the AGN cannot be excluded on energetic grounds.  It is
of great interest therefore to investigate whether strong millimeter
emission is also found in high-redshift QSOs which are optically much
fainter than those observed so far. In this respect, RD0301 is a unique
object: it is the optically least luminous QSO known at $z>4$, and it is one
of the highest redshift quasars ever detected.

Searching for high-redshift dust emission is of great interest also
because for them there was little time to produce the observed dust. At
redshift 5.5, the universe was about 1~Gyr old, a time not much longer
than the dynamical timescale of typical galaxies.  Observing large
amounts of dust at such redshifts thus sets back the first epochs of
vigorous star formation irrespective of whether the observed dust was
heated by an AGN or a starburst.

\section{Observations and results}

Our observations of RD0301 were made during the winters of 2000 and 2001
with the 37-channel {\it Max-Planck Millimeter Bolometer} array (MAMBO;
Kreysa et al.  \cite{kre99a}, \cite{kre99b}) at the IRAM 30-meter
telescope on Pico Veleta, Spain.  MAMBO is sensitive between 190 and
315~GHz, with half-power sensitivity limits at 210 and 290~GHz, and an
effective bandwidth center for steep thermal spectra at $\sim 250$~GHz.
The bolometer feed horns are matched to the telescope FHWM beam at
1.2~mm of $10\farcs7$. They are arranged in an hexagonal pattern with a
beam separation of 22$^{\prime\prime}$.  The source was observed with
the array's central channel, using the ``on-off'' observing mode, in
which the telescope secondary mirror chops in azimuth by 30 to
50$^{\prime\prime}$ at a rate of 2$\,$Hz. For on-off observations the
target is positioned on the array's central bolometer, and after every
10 seconds of integration the telescope is nodded so that the previous
``off'' beam, which typically misses the other bolometer beams, becomes
the ``on'' beam. Thus for only half the on-sky observing time the target
is positioned on any bolometer.  A scan usually lasts for 12 or 16 such
10 sec sub-scans, i.e., 6 to 8 on-off cycles.

\begin{figure}[]
  \psfig{figure={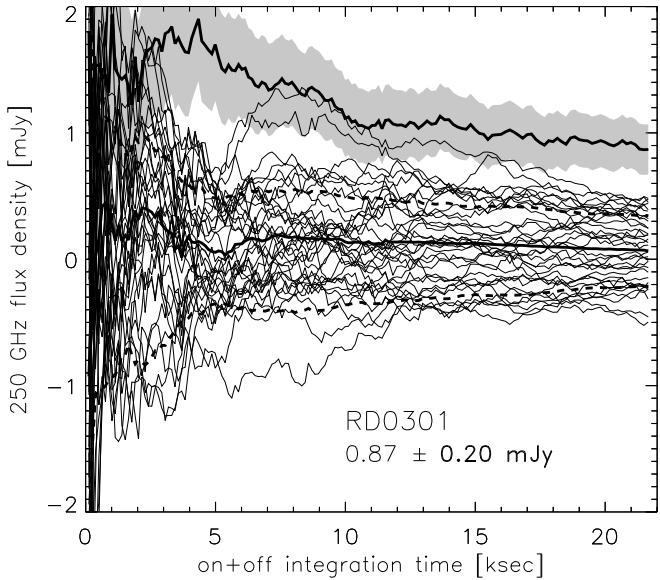},width=8.5cm} 
  \caption{Time-averaged, opacity corrected, and sky-noise subtracted flux
    densities for 33 of the 37 MAMBO bolometers. Four disfunctional
    bolometers are ignored.  The thick line shows the time-averaged
    signal from the central, on-source bolometer, the grey-shaded
    envelope is reflects its 1$\sigma$ error. The thin solid lines are
    the average on-sky bolometer signals, and the dashed lines show the
    1$\sigma$ scatter of the average on-sky bolometer signals.  }
\label{figure1}
\end{figure}

The telescope pointing was monitored regularly on two $\sim 10$~deg
distant pointing sources, 0336$-$019 and 0221$+$067, and was in most
cases stable within $\sim 2^{\prime\prime}$. The signals from the
pointing sources were monitored for sensitivity problems.  The gain
calibration was performed using observations of Mars, Uranus, and Ceres.
We here adopt a calibration factor of 12\,500 counts per Jansky for 0.5
sec integration bins, a value which is expected to be accurate to within
20\%.  The point-source sensitivity was on average $\sim 30\rm
~mJy~s^{1/2}$, i.e., in one second of on-sky (on and off source)
integration, an rms noise level of 30 mJy is reached.

The data were analyzed using the MOPSI software package created by
Robert Zylka (see Zylka \cite{zyl98}).  Correlated sky noise was
subtracted from each channnel as the weighted mean of the signals
from the best-correlating surrounding channels.

RD0301 was observed on seventeen different nights between February 2000
and March 2001. The atmospheric conditions were mostly very good, with
zenith opacities between 0.1 and 0.2. From a total of 184 scans, each
with 120 or 180~sec on-sky exposure, we eliminated one observation (1
March 2000) of 15 scans because of a 5$^{\prime\prime}$ pointing drift.
Six other scans were eliminated throughout because they showed unusual
noise, possibly due to passing clouds. To improve the quality of the
data we furthermore ignored 10 scans at elevations below 27 degrees,
where the gain-elevation and opacity corrections become large and
uncertain.

The total on-sky integration time was 6.0 hours.  The scans were reduced
individually, and the signals were then averaged, which yields a target
flux of $0.87 \pm 0.20 \, \rm mJy$ (Fig.~\ref{figure1}).  If the scans
with elevation $<27$ deg were included, the signal rises to $0.93\pm
0.21$ mJy. When weighting the scan signals by each scans' inverse
average mean square noise level, the signal becomes $0.76 \pm 0.19 \,
\rm mJy$. The dispersion in the integrated signals of the off-target
channels is 0.26~mJy. The source signal appears stronger in the winter
1999/2000 observations, which might in part be due to the fact that many
of these observations were performed at low elevation, making them
subject to larger gain-elevation and opacity uncertainties.

The coadded data of both years show an average signal that stabilizes at
a significant level compared with the target channel's integrated noise,
and also compared with the signal of the off-target channels.  We
therefore believe that the source is detected with the confidence level
suggested by the errors quoted.

To our knowledge this is the deepest integration ever at millimeter or
submillimeter wavelengths, and the faintest thermal emission ever
detected. As a promising outlook we note that this object would be
detected by ALMA in a few seconds, with an angular resolution of less
than 0.1 arcsec.

\section{Discussion} 

\subsection{Thermal nature of the millimeter emission}

At redshift 5.5 our observing frequancy of 250 GHz corresponds to an
emitted frequency of 1625~GHz or a wavelength of 185~$\rm \mu m$.  From
a flux measurement at only one frequency we cannot distinguish whether
the emission follows a steep thermal spectrum, or constitutes the
millimeter extension of a radio synchrotron spectrum. The VLA 1.4~GHz
FIRST survey (Becker, White \& Helfand \cite{bec95}) however yields a
5$\sigma$ upper limit flux density of $\simeq$~1~mJy within
30$^{\prime\prime}$ from RD0301. Since rising non-thermal spectra are
not observed at high radio frequencies, it is unlikely that the
millimeter emission is synchrotron radiation. Carilli et al.
(\cite{car01a}, \cite{car01b}) show that for most of the high-redshift
QSOs detected at 1.2 mm warm dust is the likely source of the emission.

\subsection{FIR luminosity, dust mass, and star formation rate}

We therefore conclude that the millimeter flux of RD0301 is thermal in
nature. When matching the observed 1.2~mm flux of 0.87 mJy to a grey
body with a dust temperature of 50~K and an emissivity index $\beta=1.5$
(well adapted for high-redshift sources -- see, e.g., Benford et al.
1999), the FIR luminosity of RD0301 is $L_{\rm FIR}=4.0\times
10^{12} \, \rm L_\odot$, and the implied dust mass $\approx 10^8 \, \rm
M_\odot$. The FIR luminosity of RD0301 is comparable to its optical
luminosity, $L_{\rm opt}\approx 3.7\times 10^{12}\, \rm L_\odot$, given
by the rest frame blue magnitude $M_{\rm B} \approx -24.0$ (for a
$\Lambda$-cosmology). We here adopted the blue luminosity bolometric
correction factor of $\sim 12$ derived for the PG sample by Elvis et al.
(\cite{elv94}).
%

Both the FIR luminosity and the dust mass are comparable to those
derived for local infrared-luminous galaxies such as Arp~220 and Mrk~231
(e.g., Radford et al. \cite{rad91}).  The thermal luminosity of RD0301
is comparable also to that of the bright high-redshift starburst
galaxies found in deep optical surveys (Adelberger \& Steidel
\cite{ade00}), and of those thought to be producing the bulk of the FIR
background emission (Peacock et al. \cite{pea00}).  If the FIR
luminosity of RD0301 arose from a continuous starburst of age 10 to
100~Myr with a modified Salpeter IMF (see Omont et al.  \cite{omo01} for
details on the adopted starburst model), the implied star formation rate
would be $\approx 600 \, \rm M_\odot~yr^{-1}$.  This value again is
similar to those inferred for Arp~200 and Mrk~231.

  
\subsection{Comparison with the SDSS and DPSS QSOs}

The millimeter flux of RD0301 is typical for the optically brightest
high-redshift QSOs.  When averaging (as if we observed one single
object) the MAMBO observations of the 112 redshift 3.6 to 5.0 QSOs
observed in the Carilli et al. (\cite{car01a}) SDSS and Omont et al.
(\cite{omo01}) DPSS surveys, one obtains an average flux density of
$1.3\pm 0.05$~mJy, which implies an average luminosity of $L_{\rm FIR}
\approx 5\times 10^{12} \, \rm L_\odot$.  The dust luminosity of RD0301
is therefore close to the average dust luminosity of a sample of quasars
with $-26>M_{\rm B}>-29.5$, sources which are optically 10 to 100 times
brighter than RD0301.

\begin{figure}[]
  \psfig{figure={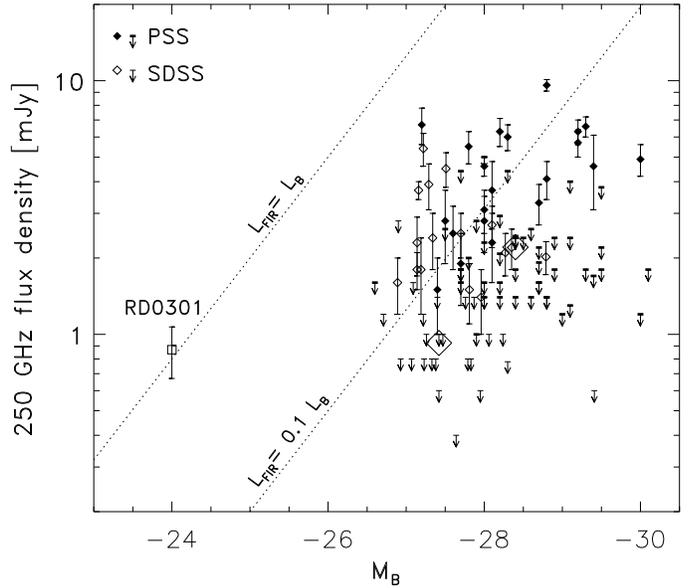},width=9cm} 
  \caption{MAMBO 1.2 mm flux density as a function of blue magnitude for
    the $z>3.6$ QSOs observed in the Carilli et al. (open symbols) and
    Omont et al. (filled symbols) surveys. Average source fluxes for the
    respective samples are indicated as large open squares, which are
    placed at the median $M_{\rm B}$ of the respective sample.  }
\label{figure2}
\end{figure}

The mean blue magnitudes of the Carilli et al. (\cite{car01a}) and Omont
et al. (\cite{omo01}) QSOs are $-27.0$ and $-27.9$ mag, respectively
(Fig.~\ref{figure2}).  Although the QSOs of these samples span a range
of about 25 in optical brightness, the DPSS QSOs studied by Omont et al.
(\cite{omo01}) are on average twice as bright in the optical as the SDSS
QSOs observed by Carilli et al. (\cite{car01a}).  Similarly, the average
1.2~mm flux of the QSOs in the respective samples differs by a factor
two, $0.92$ mJy versus $2.1$~mJy (large squares in Fig.~\ref{figure2}).
However, the significance of a possible correlation between the average
optical and millimeter brightness of high $z$ QSOs remains low since the
scatter in both quantities is much larger than the difference between
the samples' averages (see Omont et al. [\cite{omo02}] for a detailed
statistical discussion).

RD0301 does not follow the possible trend of increasing millimeter flux
with optical luminosity seen in the QSO millimeter surveys, as Fig.
\ref{figure2} well illustrates.  Further (sub)millimeter observations of
optically faint ($M_{\rm B}>-26$) high-redshift QSOs are needed to
clarify the relation between their thermal and optical luminosities.  A
lack of such a correlation would support the idea that the dust heating
is caused by young, massive stars, and not by the AGN.  This hypothesis
is also supported by the detection of CO emission in a number of
high-redshift QSOs (Guilloteau \cite{gui01}; Cox et al. \cite{cox02}),
and by the lack of strong radio emission from most of them (Carilli et
al. \cite{car01b}).

Alternatively, unification models of AGN might not necessarily predict a
strong correlation between the orientation-dependent optical emission
and the isotropic FIR flux from the AGN.
But still the AGNs' FIR flux should correlate well with its hard X-ray
emission, which does not suffer much from extinction. Millimeter-studies
of high-redshift, hard X-ray sources would be most useful to address
this interesting issue (e.g. Page et al. \cite{pag01}).  SIRTF
observations of mid-IR PAH features should also be able to clarify the
origin of the FIR emission: a strong correlation between PAH and FIR
emission supports the starburst origin of the FIR luminosity (Genzel et
al. \cite{gen98}).

\acknowledgements 

We wish to thank E.~Kreysa, L.~Reichertz and the MPIfR bolometer group
for building MAMBO, and R.~Zylka for providing the excellent MOPSI data
reduction package. Thanks also to Alain Omont, Chris Carilli, Hauke Voss,
Alexandre Beelen, and the IRAM staff for help with the observations and
interpretation. The constructive comments and criticism from Daniel Stern
are much appreciated.  IRAM is supported by INSU/CNRS (France), MPG
(Germany), and IGN (Spain).

\end{document}